\documentstyle[color,aps,prb,psfig,epsfig]{revtex}
\begin{document}
%\draft
%\input{mathmac}
% Formula MACROS enter here #####################################################
\def\bdm{\begin{displaymath}}
\def\edm{\end{displaymath}}
\def\nn{\nonumber}
\def\bc{\begin{center}}
\def\ec{\end{center}}
\def\be{\begin{equation}}
\def\ee{\end{equation}}
\def\tcb{\textcolor{blue}}
\def\tcbl{\textcolor{black}}
\def\tcg{\textcolor{green}}
\def\tcr{\textcolor{red}}
\def\tcgr{\textcolor{grey}}
\def\va{{\bf a}}
\def\vA{{\bf A}}
\def\vb{{\bf b}}
\def\vB{{\bf B}}
\def\vb{{\bf b}}
\def\vc{{\bf c}}
\def\vC{{\bf C}}
\def\vd{{\bf d}}
\def\hvd{\hat\vd}
\def\vD{{\bf D}}
\def\ve{{\bf e}}
\def\hve{\hat\ve}
\def\vE{{\bf E}}
\def\vf{{\bf f}}
\def\vF{{\bf F}}
\def\vg{{\bf g}}
\def\vG{{\bf G}}
\def\vh{{\bf h}}
\def\vH{{\bf H}}
\def\vi{{\bf i}}
\def\vI{{\bf I}}
\def\vj{{\bf j}}
\def\vJ{{\bf J}}
\def\vk{{\bf k}}
\def\hvk{\hat\vk}
\def\vK{{\bf K}}
\def\vl{{\bf l}}
\def\vL{{\bf L}}
\def\vLambda{{\bf\Lambda}}
\def\vm{{\bf m}}
\def\vM{{\bf M}}
\def\vn{{\bf n}}
\def\hvn{\hat\vn}
\def\vN{{\bf N}}
\def\vone{{\bf 1}}
\def\vp{{\bf p}}
\def\hvp{\hat\vp}
\def\vP{{\bf P}}
\def\vq{{\bf q}}
\def\vQ{{\bf Q}}
\def\vr{{\bf r}}
\def\vR{{\bf R}}
\def\vs{{\bf s}}
\def\vS{{\bf S}}
\def\vt{{\bf t}}
\def\vT{{\bf T}}
\def\vu{{\bf u}}
\def\vU{{\bf U}}
\def\vv{{\bf v}}
\def\vV{{\bf V}}
\def\vw{{\bf w}}
\def\vW{{\bf W}}
\def\vx{{\bf x}}
\def\vX{{\bf X}}
\def\vy{{\bf y}}
\def\vY{{\bf Y}}
\def\vz{{\bf z}}
\def\v0{{\bf 0}}
\def\hvz{\hat\vz}
\def\vZ{{\bf Z}}
\def\vtau{{\bf \tau}}
\def\e{{\rm e}}
\def\kB{k_{\rm B}}
\def\kF{k_{\rm F}}
\def\EF{E_{\rm F}}
\def\NF{N_{\rm F}}
\def\pF{p_{\rm F}}
\def\Tc{T_{\rm c}}
\def\vvF{v_{\rm F}}
\def\vna{{\bf\nabla}}
\def\vPi{{\bf\Pi}}
%
% End of formula MACRO list ######################################################
%
\twocolumn[\hsize\textwidth\columnwidth\hsize
\csname@twocolumnfalse\endcsname
\title{Electronic Raman response in anisotropic metals}
\author{Dietrich Einzel$^1$ and Dirk Manske$^2$}
\address{$^1$ Walther--Meissner--Institut, Bayerische Akademie der
Wissenschaften, D--85748 Garching, FRG}
\address{$^2$Max--Planck--Institut f\"ur Festk\"orperforschung,
Heisenbergstrasse 1, D--70569 Stuttgart, FRG}
\date{\today}
\maketitle
\begin{abstract}
Using a generalized response theory we derive the electronic Raman
response function for metals with anisotropic relaxation rates.
The calculations account for the long--range Coulomb interaction
and treat the collision operator within a charge conserving
relaxation time approximation. We extend earlier treatments to
finite wavenumbers ($|{\bf q}|\ll k_{\rm F}$) and incorporate
inelastic electron--electron scattering besides elastic impurity
scattering. Moreover we generalize the Lindhard density response
function to the Raman case. Numerical results for the
quasiparticle scattering rate and the Raman response function for
cuprate superconductors are presented.
\end{abstract}
]

\noindent {\bf Introduction} \\
After the discovery of high-$\Tc$ superconductors, something of a
stir has been caused by the possibility of investigating these
systems by Raman scattering experiments. Through its polarization
dependence this spectroscopic method allows for detecting
anisotropies in quantities like the superconducting gap and the
scattering rates characterizing the transport properties. While
the gap anisotropies in high-$\Tc$ cuprates, as seen in Raman
experiments, are theoretically well understood in terms of
$d_{y^2-y^2}$ pairing\cite{DANDE}, the important quasiparticle
collision effects are much less studied. Early attempts included
constant scattering rates\cite{FANDK}, elastic scattering
processes \cite{ZANDC,EANDS} or described inelastic scattering
processes within the Nearly Antiferromagnetic Fermi Liquid (NAFL)
model\cite{DANDK}.

In this paper we investigate the general situation of a normal
metal by developing a theory of the electronic Raman effect at
finite wavenumbers $\vq\not=0$ within the RPA response theory and
generalize it to include collision effects. A superposition of
elastic and inelastic scattering processes is considered with
$\vk$--dependent relaxation rates and deviations from
Matthiessen's rule are studied. Having in mind quasi 2D-systems
like high-$T_c$ cuprate superconductors, we present a numerical
analysis of the inelastic scattering rates and the Raman response
functions employing the FLuctuation EXchange (FLEX) approximation
which treats the spin fluctuation--limited transport and
Cooper-pairing on the same level.\\%%[1ex]
\noindent {\bf Transport theory and generalized Raman\\
response: normal state}\\
We consider a normal metal in which the electronic states are
characterized by a momentum $\hbar{\bf k}$, an energy dispersion
$\epsilon_\vk=\xi_\vk+\mu$, (we set the lattice constant to unity)
\begin{equation}
\epsilon_{\bf k}= t\left[ -2\cos(k_x) - 2\cos(k_y) +
4B\cos(k_x)\cos(k_y) \right] \label{eq:epsilon} \, ,
\end{equation}
with $\mu$ being the Fermi energy, the group velocity ${\bf
v}_\vk=(1/\hbar)\vna_\vk\epsilon_\vk$, an inverse effective mass
tensor $M_{i j}^{-1}({\bf
k})=\partial^2\epsilon_\vk/\hbar^2\partial k_i\partial k_j$, and
an equilibrium Fermi distribution $n_\vk$ with derivative
$\varphi_\vk=-\partial n_\vk/\partial\xi_\vk$. Next we focus on an
external perturbation appropriate for a treatment of the
electronic Raman response within the effective mass
approximation:\cite{wolff,cardona,manskefay}
\be U_\vk^{\rm\ ext}=\underbrace{m\hat{\bf e}^{\rm S}\cdot
\vM^{-1}(\vk)\cdot\hat{\bf e}^{\rm I}}_{\gamma_\vk}
\cdot\underbrace{\frac{e^2}{mc^2}|\vA^{\rm I}||\vA^{\rm
}|}_{u_\gamma^{\rm ext}}\, . \label{eq:u}
 \ee
Here, $\hat{\bf e}^{\rm I}$ and $\hat{\bf e}^{\rm S}$ are the unit
vectors of the incident and scattered light, respectively, and
$\vA$ denotes the vector potential. Then, the response of the
electronic system to this perturbation is described within the the
quasiclassical limit of the kinetic equation:
\be \omega\delta n_\vk-\vq\cdot\vv_\vk h_\vk=i\sum_{\nu={\rm
e},{\rm i}}\delta I_\vk^\nu\, , \ee
where we use the definition
\begin{equation}
h_\vk = \delta n_\vk+\varphi_\vk \left[ U_\vk^{\rm
ext}+V(\vq)\delta n_1 \right]
\end{equation}
and $V(\vq)=4\pi e^2/\vq^2$ is the Fourier transform of the long
range Coulomb interaction. Physical observables are the
generalized response functions
\be \delta n_a=\sum_{\vp\sigma}a_\vp\delta n_\vp \, ,
\label{apbp}\ee
where $a_\vp$ is the vertex which describes the coupling of
$\delta n_a$ to the external perturbation potential.
The collision integrals for elastic ($\nu=$e) and inelastic
($\nu=$i) scattering have the form (conserving relaxation time
approximation)
\begin{eqnarray}
\delta I_\vk^\nu&=&-\Gamma_\vk^\nu
h_\vk+\sum_{\vp\sigma}C_{\vk\vp}^\nu h_\vp \quad ,\nn \\
C_{\vk\vp}^\nu&\approx&\varphi_\vk\sum_{b}\lambda_b^\nu
\frac{b_\vk\Gamma_\vk^\nu
b_\vp\Gamma_\vp^\nu}{\sum_{\vp\sigma}\varphi_\vp
b_\vp^2\Gamma_\vp^\nu} \quad .
\end{eqnarray}
The scattering parameters $\lambda_b^\nu$ allow a classification
of the macroscopic moments $\delta n_b$ (defined through Eq.
(\ref{apbp})) into conserved ($\lambda_b^\nu=1$) and nonconserved
($\lambda_b^\nu<1$) quantities. In what follows we will, for the
sake of simplicity, restrict ourselves to the case of charge
conservation, i.e. $\lambda_1^\nu=1$, $\lambda_b^\nu=0$ $\forall
b\not=1$. While the elastic scattering rate $\Gamma_\vk^i$ is
constant, the inelastic part of the scattering rate $\Gamma_\vk^i$
has a strong frequency dependence that reflects characteristic
lifetime effects, which we discuss below.

\begin{figure}[t]
%%\centerline{\psfig{file=hh01a.eps,width=8.5cm,angle=-90}}
\includegraphics[bb= 85 14 498 633,width=5.5cm,angle=-90,clip]{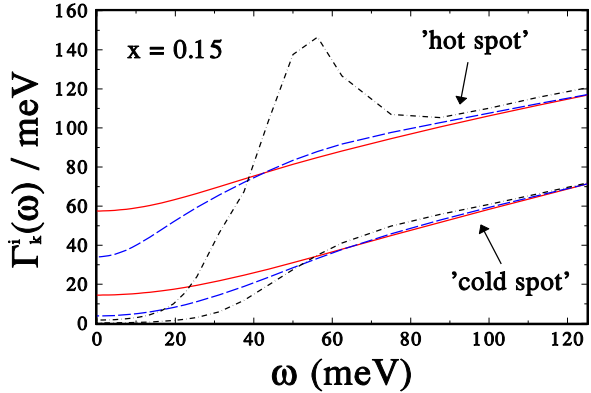}
\vspace*{1ex} \caption{Calculated inelastic scattering rate
$\Gamma_\vk^i$ versus $\omega$ for the one-band Hubbard model
using $U=4t$ at optimal doping ($x=0.15$) for temperatures
$T=2T_{\rm c}$ (solid lines), $T=1.05T_{\rm c}$ (dashed lines),
and $T=0.5T_{\rm c}$ (dashed--dotted lines).\vspace*{0ex}}
\label{Fig1}
\end{figure}

In Fig. \ref{Fig1} we show the inelastic scattering rate
$\Gamma_\vk^i(\omega)$ for hole-doped high-$T_c$ cuprates at
optimal doping ($x=0.15$) for various temperatures using the FLEX
approximation \cite{Manskeetal03,Bickers,DahmTewordt,Bennemann}.
The upper curves correspond to $\vk$--directions near $(\pi,0)$
('hot spots'), the lower ones refer to $\vk$ near the diagonal
('cold spots'). For $\omega \to 0$ we find that the inelastic
scattering rate at the hot spot is almost three times larger than
at the cold spot (solid line). Physically speaking, the scattering
rates $\Gamma_\vk^i$ can be understood in terms of scattering of
quasiparticles on spin fluctuations (paramagnons) which are
enhanced near the hot spots. In the normal state we also find at
small frequencies that the scattering rates decrease with
decreasing temperature. For $T<T_{\rm c}$ a rearrangement of
spectral weight occurs reflecting the $\omega$-dependence of the
superconducting $d_{x^2-y^2}$-wave gap $\Delta(\omega)$ which is
calculated self-consistently \cite{manskeprl}. As expected, the
maximum is seen at $\omega\simeq 3\Delta_0/\hbar$.  In the
high-frequency limit, $\Gamma_\vk^i$ varies linearly with $\omega$
for all $\vk$--directions. This is in accordance with both the
Marginal (MFL)\cite{varma} and Nested (NFL)\cite{ruvalds} Fermi
liquid picture.
\vspace*{-1ex}

\begin{figure}[t]
%%\centerline{\psfig{file=hh02a.eps,width=8.5cm,angle=-90}}
\includegraphics[bb= 85 14 498 633,width=5.5cm,angle=-90,clip]{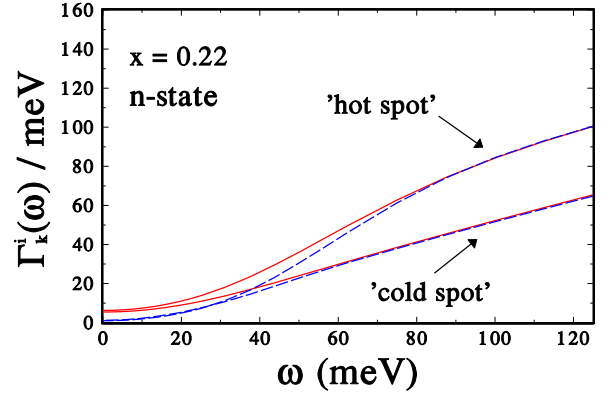}
\vspace*{1ex} \caption{Inelastic scattering rate $\Gamma_\vk^i$
versus $\omega$ for the overdoped case ($x=0.22$) using the same
notation as in Fig. \protect{\ref{Fig1}}.} \label{Fig2}
\end{figure}

Let us turn to the overdoped case ($x=0.22$) where we focus on the
normal state. Thus, we show in Fig. \ref{Fig2} $\Gamma_\vk^i$ for
the same temperatures and directions as in Fig. \ref{Fig1}. Most
importantly, the scattering rates become less anisotropic at small
frequencies.  This is in agreement with Raman scattering
experiments where the static scattering rate is extracted for
different high-$T_c$ cuprates for different polarisations as a
function of the doping concentration \cite{opel,venturini}.
Generally speaking, the anisotropy of the scattering rate reflects
the anisotropy of the spin fluctuations. Those become less
pronounced and more isotropic in the overdoped regime.  We still
find a reduction of $\Gamma_\vk^i$ with decreasing temperature and
the linear high--frequency behavior remains.

Next, we turn to the generalized theory of Raman scattering. The
full response function ${\cal L}_{ab}$ for general vertices $a$
and $b$ reads within RPA (see (Eq. \ref{eq:u}))
\vspace*{-1ex}
\be {\cal L}_{ab}=\frac{\delta n_a}{\delta u^{\rm ext}_b}=
\underbrace{L_{ab}-\frac{L_{a1}L_{1b}}{L_{11}}}_{\rm TRANSVERSAL}
+ \underbrace{\frac{L_{a1}L_{1b}}{L_{11}}\, \frac{1}{1-I
L_{11}}}_{\rm LONGITUDINAL}   \, . \ee
$I$ denotes the corresponding irreducible interaction in a
symbolic notation. Note that the decomposition into transverse and
longitudinal parts is a general structural property of the respose
formalism. For the special case of Raman scattering ($a,b =
\gamma_{\bf k}$) one has to consider the Lindhard response and
$(1-IL_{11})$ is the dielectric function $\epsilon$. Then, the
full Raman response is of the form
\begin{eqnarray}
L_{\gamma\gamma}(\vq,\omega)&=& \frac{\delta
n_\gamma(\vq,\omega)}{u^{\rm ext}_\gamma(\vq,\omega)} \label{fatL}
=M_{\gamma\gamma}^*-\frac{M_{\gamma1}^{*
2}}{M_{11}^*} \left(1-\frac{1}{\epsilon}\right) \nn \\
&-&\Xi_{\gamma\gamma}^*+\frac{\Xi_{\gamma1}^{* 2}}{\Xi_{11}^*}
+\frac{\Xi_{11}^*\zeta^*_{\gamma\gamma}}{\epsilon}
+O\left(\vq^2,\frac{1}{\epsilon}\right) \, .
\end{eqnarray}
Here, $M_{ab}^*$ is the generalization of the Lindhard function
due to the inclusion of collision effects:
\be M_{ab}^*=\frac{\vq\cdot\left(\vT_{ab}^{(0)*}+
\zeta_{ab}^*\vT_{11}^{(1)*}-\vT_{ab}^{(1)*}\right)
\cdot\vq}{i\omega-\vq\cdot\vD_{11}^{(1)*}\cdot\vq}\, .
\label{eq:mab} \ee
$\vT_{ab}^{(\mu)*}$ are generalizations of the electronic
conductivity to general vertices $a$, $b$:
\begin{eqnarray}
\vT_{ab}^{(\mu)*}&=& \sum_{\vp\sigma}\left(-\frac{\partial
n_\vp}{\partial\xi_\vp}\right)
f_\vp\frac{a_\vp\vv_\vp:b_\vp\vv_\vp}{-i\omega+\Gamma_\vp^*}
\left(\frac{i\Gamma_\vp^*}{\omega+i\Gamma_\vp^*}\right)^\mu \, ,\\
f_\vp&=&=\frac{(\omega+i\Gamma_\vp^*)^2}{\left(\omega+
i\Gamma_\vp^*\right)^2-(\vq\cdot\vv_\vp)^2}\, , \,
\Gamma_\vp^*=\Gamma_\vp^{\rm e}+\Gamma_\vp^{\rm i}\nn
\end{eqnarray}
The quantities $\vD_{ab}^{(\mu)*}$ are generalized diffusion
tensors
\begin{displaymath}
\vD_{ab}^{(\mu)*} = \frac{\vT_{ab}^{(\mu+1)*}}{N_{ab}^*} \, , \,
N_{ab}^* =  \sum_{\vp\sigma}\left(-\frac{\partial
n_\vp}{\partial\xi_\vp}
\right)\frac{i\Gamma_\vp^*}{\omega+i\Gamma_\vp^*}a_\vp b_\vp\, .
\end{displaymath}
$\Xi_{ab}^*$ are the collision--limited Raman response functions
which have a finite $\vq\to 0$ limit:
\be \Xi_{ab}^*=\sum_{\vp\sigma}\left(-\frac{\partial
n_\vp}{\partial\xi_\vp}\right)
f_\vp\frac{i\Gamma_\vp^*}{\omega+i\Gamma_\vp^*}a_\vp b_\vp \ee
and the objects $\zeta_{ab}^*$ describe the mixing of elastic and
inelastic scattering processes and hence deviations from
Matthiessen's rule within the conserving relaxation time
approximation:
\begin{eqnarray}
\zeta_{ab}^*&=& \frac{\Xi_{a1}^*\Xi_{1b}^*}{\Xi_{11}^{* 2}}
+\frac{\omega}{\omega+i\gamma^*} \frac{\Xi_{11}^{\rm
e}\Xi_{11}^{\rm i}}{\Xi_{11}^{*2}} \nn\\
&\times&\left(\frac{\Xi_{a1}^{\rm e}}{\Xi_{11}^{\rm
e}}-\frac{\Xi_{a1}^{\rm i}}{\Xi_{11}^{\rm i}}\right)
\left(\frac{\Xi_{b1}^{\rm e}}{\Xi_{11}^{\rm
e}}-\frac{\Xi_{b1}^{\rm i}}{\Xi_{11}^{\rm i}}\right)
+O(\vq^2)\\
\gamma^*&=&\frac{\Xi_{11}^*}{\Xi_{11}^{\rm e}\Xi_{11}^{\rm i}}
\sum_{\vp\sigma}\left(-\frac{\partial
n_\vp}{\partial\xi_\vp}\right)
f_\vp\frac{i\Gamma_\vp^{\rm e}\Gamma_\vp^{\rm i}}{\omega+i\Gamma_\vp^*}\nn\\
\Xi_{ab}^\nu&=&\sum_{\vp\sigma}\left(-\frac{\partial
n_\vp}{\partial\xi_\vp}\right)
f_\vp\frac{i\Gamma_\vp^\nu}{\omega+i\Gamma_\vp^*} a_\vp b_\vp \ \
\ ; \ \ \ \nu={\rm e},{\rm i}\nn
\end{eqnarray}
Finally,
$\epsilon=\epsilon(\vq,\omega)=1-V(\vq)M_{11}^*(\vq,\omega)$ is
the dielectric function of the electronic system mentioned
earlier. It has thus been shown that the mixing terms
$\propto\zeta_{ab}^*$ in Eq. (\ref{fatL}) of the two separate
scattering mechanisms occur in the Lindhard function
$M_{ab}^*(\vq,\omega)=O(\vq^2)$ and in terms which are screened
$\propto\epsilon^{-1}$ by the long range Coulomb interaction. For
practical applications of the result shown in Eq. (\ref{fatL}) to
the cuprate systems, say, where $\vq\to 0$ and $\epsilon\approx
10^4$ may be assumed, one may use the $\vq\to 0,
\epsilon\to\infty$ limit of $L_{\gamma\gamma}$ and the
contributions from mixing $\propto\zeta_{ab}^*$ are hence
irrelevant. Therefore, the scattering mechanisms can be linearly
combined, leading to the exclusive occurrence of
$\Gamma_\vp^*=\Gamma_\vp^{\rm e}+\Gamma_\vp^{\rm i}$ in the Raman
response function. Note that the Raman response function $\mbox{Im
}{\cal L}_{\gamma\gamma}$ in the limit ${\bf q} \to 0$ reduces to
Eq. (12) of Ref. \onlinecite{ZANDC} where
$\left(\tau_L^{*}\right)^{-1}$ corresponds to an average of
$\Gamma_\vp^*$. Hence, this scattering rate can be interpreted as
the width of the lineshape of Fig. 3 in Ref. \onlinecite{ZANDC}.
Next we focus on the numerical analysis of $\mbox{Im }{\cal
L}_{\gamma\gamma}$ for a spin-fluctuation based model particularly
for the inelastic scattering rate $\Gamma_\vk^{\rm i}$.\\%%[1ex]
\noindent {\bf Numerical calculation of the Raman response} \\
Using normal and anomalous Green's functions, $G$ and $F$, we
calculate the full Raman response function via
\begin{eqnarray}
\mbox{Im }{\cal L}_{\gamma\gamma}({\bf q}=0,\omega) & = &
\pi\int_{-\infty}^{\infty}d\omega' \left[f(\omega') -
f(\omega'+\omega)\right]
\nonumber\\
& \hspace*{-2cm}\times & \hspace*{-1cm}\sum_{\bf
k}\tilde\gamma({\bf k},\omega',\omega) \left[N({\bf
k},\omega'+\omega)N({\bf k},\omega') \right.
\nonumber\\
& \hspace*{-2cm}- & \hspace*{-1cm}\left. A({\bf
k},\omega'+\omega)A({\bf k},\omega')\right] \gamma({\bf k})\ \ ,
\label{eq:ramanresponse}
\end{eqnarray}
where, $N=G''$ and $A=F''$ are the spectral functions (i.e. the
${\bf k}$-resolved density of states) of the corresponding
quasiparticles calculated within the FLEX approximation for the
one-band Hubbard model \cite{Manskeetal03,comment}. The quantity
$\tilde\gamma(\vk,\omega^\prime,\omega)$ denotes the renormalized
Raman vertex that will be specified below. The bare Raman vertices
for the different polarization symmetries considered here,
$B_{1g}$, $B_{2g}$, read within the effective mass approximation:
\begin{displaymath}
\gamma_{B_{1g}} = t\left[\cos(k_x) - \cos(k_y)\right] ,
\gamma_{B_{2g}} = 4tB\sin(k_x)\sin(k_y) .
\end{displaymath}
Here, $t$ is the nearest neighbor and $t'=-Bt$ (with $B=0.45$) is
the next-nearest neighbor hopping energy of the tight-binding band
introduced in Eq. (\ref{eq:epsilon}). Thus Raman scattering in
$B_{1g}$ symmetry mainly probes the 'hot regions' while $B_{2g}$
symmetry probes the 'cold regions' of the Brillouin zone.

It has been shown earlier that the FLEX approximation yields a
phase diagram $T(x)$ for cuprates which is in fair agreement with
experiment. One finds a $d_{x^2-y^2}$-wave superconducting order
parameter and all characteristic temperature scales
\cite{manskedahm01}.  Recently it has also been demonstrated that
the resonance peak below $T_c$ in neutron scattering data and
angle-resolved photoemission (ARPES) experiments which measure the
spectral density entering in Eq.(\ref{eq:ramanresponse}) can be
well described \cite{manskeins01,manskeprl}.

\begin{figure}[t]
%%\centerline{\psfig{file=hh03a.eps,width=8.5cm,angle=-90}}
\includegraphics[bb= 83 11 500 627,width=5.5cm,angle=-90,clip]{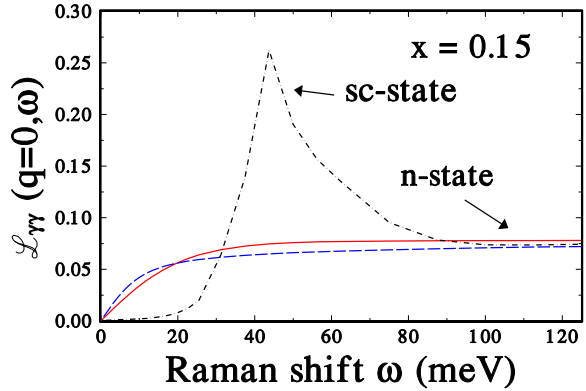}
\vspace*{1ex} \caption{Full Raman response function ${\cal
L}_{\gamma\gamma}(\vq\to 0,\omega)$ versus $\omega$ at optimal
doping ($x=0.15$) for $B_{1g}$ polarization for the same
temperatures as in Fig. \protect{\ref{Fig1}}.} \label{Fig3}
\end{figure}

While no vertex corrections are considered using the kinetic
equation approach, they become important if the FLEX approximation
is employed. The main consequence is that the Raman vertex becomes
frequency dependent reflecting the retardation effects related to
the Cooper-pairing mechanism. In order to calculate the vertex
function $\gamma_\nu$ we employ the Nambu notation ($\mu,\nu =
0,1,2,3$) \cite{Nambu}
\begin{equation}
P_{\mu\nu}(q)= -\sum_{k} \frac{1}{2}\mbox{Tr}\left[ \gamma_{\mu}
G(k+q)\tilde\gamma_{\nu} G(k)\right] \label{A1}
\end{equation}
with $q\equiv {\bf q},i\nu_m; \, k\equiv {\bf k},i\omega_n; \,
\sum_{k} = T\sum_{i\omega_n}\sum_{{\bf k}}$, and satisfy the Ward
identity
%
%%\begin{equation}
$\sum_{\mu}q_{\mu}\tilde\gamma_{\mu}(k+q,k) = \tau_3G^{-1}(k) -
G^{-1}(k+q)\tau_3$.
%%\quad . \label{A6}
%%\end{equation}
%
The ladder approximation for the vertex function yields
\begin{eqnarray}
\lefteqn{\tilde\gamma_{\mu}(k+q,k) = \gamma_{\mu}(k+q,k)
}\nonumber\\
& + &
\sum_{k'}\left[\tau_0G(k'+q)\tilde\gamma_{\mu}(k'+q,k')G(k')\tau_0
P_s(k-k')\right.\nonumber\\
& + & \left.
%\hspace{0.6cm}
\tau_3G(k'+q)\tilde\gamma_{\mu}(k'+q,k')G(k')\tau_3P_c(k-k')\right]
. \label{A5}
\end{eqnarray}
$P_s$ and $P_c$ refer to the spectral function of the spin and
charge susceptibility within RPA, respectively, that are defined
in Ref. \onlinecite{Manskeetal03}. Finally, the full Raman vertex
$\tilde\gamma$ satisfies the following integral equation:
\begin{eqnarray}
\lefteqn{\tilde\gamma(k+Q,k) = \gamma({\bf k})
}\nonumber\\
& + & \sum_q \left[ P_s(q) + P_c(q)\right]\frac{1}{2}\mbox{Tr}
\left[\tau_3G(k+q+Q)\tau_3G(k+q)\right]
\nonumber\\
& \times & \hspace{0.7cm} \tilde\gamma(k+q+Q,k+q) \label{A9}
\end{eqnarray}
and can be derived from Eq. (\ref{A1}) by replacing $\gamma_0$ and
$\tilde\gamma_0$ by $\gamma\tau_3$ and $\tilde\gamma\tau_3$.\\%%[1ex]
\noindent {\bf Results for ${\cal L}_{\gamma\gamma}$ and discussion}\\
In Fig. \ref{Fig3} we show our results for the full Raman response
function Eq. (\ref{fatL}), $\cal{L}_{\gamma\gamma}$, for the
$B_{1g}$ polarization, which is proportional to the measured Raman
intensity, at optimal doping. Both a linear increase and a
continuum are found in the limit of low and high frequencies,
respectively. This is a direct consequence of the inelastic
scattering rate discussed in Fig. \ref{Fig1}. In particular, we
find an increasing initial slope with decreasing temperature.
Below $T_{\rm c}$ we obtain the usual pair--breaking feature
\vspace*{-1ex}
\noindent accompanied by a suppression of spectral weight for
$\omega\leq 2\Delta_0/\hbar$. For low frequencies we also find a
power law ${\cal L}_{\gamma\gamma} \propto (\omega/2\Delta_0)^3$
that is charcteristic for a $d_{x^2-y^2}$-wave gap in the clean
limit. Note that the pair breaking peak is finite and renormalized
by inelastic quasiparticle scattering processes.

\begin{figure}[t]
%%\centerline{\psfig{file=hh04a.eps,width=8.5cm,angle=-90}}
\includegraphics[bb= 83 11 500 627,width=5.5cm,angle=-90,clip]{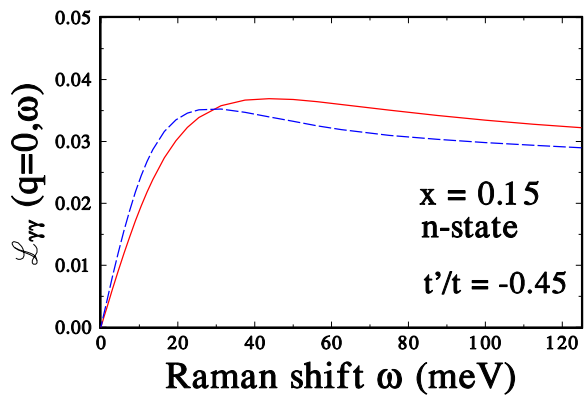}
\vspace*{1ex} \caption{Full Raman response function ${\cal
L}_{\gamma\gamma}(\vq\to 0,\omega)$ versus $\omega$ at optimal
doping ($x=0.15$) for $B_{2g}$ polarization for the same
temperatures as in Fig. \protect{\ref{Fig1}}.} \label{Fig4}
\end{figure}

Figure \ref{Fig4} shows the Raman intensity for $B_{2g}$
\vspace*{-1ex}
\noindent polarization in the normal state. We again find the
high--frequency continuum and an increasing slope of the Raman
response function with decreasing temperature. This agrees with
the NAFL picture if vertex corrections are considered
\cite{DANDK}.

\noindent
{\bf Summary}\\
In summary we have reconsidered the electronic Raman response with
special emphasis on the role of quasiparticle scattering
processes. In the normal state a superposition of elastic and
inelastic scattering events described by $\vk$--dependent
scattering rates leads to a deviation from Matthiessen's rule for
the corresponding Raman response functions. These deviations,
however, are found to be eliminated by the presence of the
long-range Coulomb interaction. We have furthermore studied in
detail, the anisotropy  of the inelastic scattering rate
$\Gamma_\vk^i$ at optimal doping, and its decrease at higher
doping levels towards the overdoped case. The resultant electronic
Raman response functions that include the effects of inelastic
scattering involving spin fluctuations reproduce the behavior of
the normal state characteristic of the phenomenological MFL
picture that has been studied in earlier
publications\cite{RUDIXY}. In the superconducting state the
inelastic scattering rate becomes gapped due to a frequency
dependent $d_{x^2-y^2}$-wave gap, calculated self-consistently,
yielding the characteristic shape of the Raman
response.\cite{DANDE}\\
\noindent {\bf Acknowledgements}\\
D.M. thanks the Walther-Meissner-Institute for hospitality and
INTAS (project 01-0654) for financial support.

\begin{thebibliography}{10}
\bibitem{comment} Note that the NAFL approach of Ref.
\protect\onlinecite{DANDK} is restricted to the normal state and
thus no spectral function $A$ occurs.
\end{thebibliography}
\end{document}